# scientific data

**OPEN**

**DATA DESCRIPTOR**

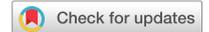

# A multimodal sensor dataset for continuous stress detection of nurses in a hospital

Seyedmajid Hosseini[1], Raju Gottumukkala[1] ✉, Satya Katragadda[1], Ravi Teja Bhupatiraju[1], Ziad Ashkar[1], Christoph W. Borst[1] & Kenneth Cochran[1,2]

Advances in wearable technologies provide the opportunity to monitor many physiological variables continuously. Stress detection has gained increased attention in recent years, mainly because early stress detection can help individuals better manage health to minimize the negative impacts of long-term stress exposure. This paper provides a unique stress detection dataset created in a natural working environment in a hospital. This dataset is a collection of biometric data of nurses during the COVID-19 outbreak. Studying stress in a work environment is complex due to many social, cultural, and psychological factors in dealing with stressful conditions. Therefore, we captured both the physiological data and associated context pertaining to the stress events. We monitored specific physiological variables such as electrodermal activity, Heart Rate, and skin temperature of the nurse subjects. A periodic smartphone-administered survey also captured the contributing factors for the detected stress events. A database containing the signals, stress events, and survey responses is publicly available on Dryad.

## Background & Summary

Prolonged exposure to stress factors such as high workload, lack of autonomy, and long hours can negatively impact employee health. Many studies point out that prolonged exposure to stress leads to chronic conditions such as obesity[1] or hypertension[2], which may exacerbate conditions such as type-II diabetes[3]. Monitoring and understanding stress in workplaces is important, especially in professions with increased exposure to stress, often leading to burnout and increased turn over[4].

This dataset provides physiological stress signals from a nursing population working in real-world hospital settings during the COVID-19 outbreak. The primary motivation to create a wearable biometric nursing stress dataset is to help advance research into understanding and improving the emotional health of nurses in a naturalistic environment through the development of early detection of work-related stress detection algorithms.

Our work was inspired by several previous works on wearables to monitor physiological signals related to stress. The AffectiveRoad dataset[5] used Empatica E4 and Zephyr Bioharness 3.0 to study the effect of driving conditions on the stress levels of 10 drivers. This study was conducted with drivers taking a 1 hour 26 minute driving test. The WESAD data-set[6] used RespiBAN and Empatica E4 to study the stress of 15 students watching a movie and taking a TSST test[7]. The SWELL dataset[8] used Mobi, uLog, video, and Kinect to study stress and the associated postures and facial expressions of 25 students. Finally, MDPSD (multi-modal dataset for psychological stress detection)[9] collected a comprehensive multimodal stress detection dataset on university students using electrodermal activity of skin (EDA) and photoplethysmography (PPG) signals while performing different tests (e.g., Trier Social Stress Test TSST[10] and color-word tests[11]).

TILES-2018 dataset from Mundnich *et al.*[12] is a multi-sensor dataset that uses a battery of surveys to cover personality traits, behavioral states, job performance, and well-being over time. Compared to TILES-2018, our dataset is much smaller in scope. Our dataset was generated during COVID-19, and the stress events we captured were linked to stress contributors. Tesserae[13–19], is a large multi-university project that studied various aspects of the workplace performance of information workers using wearables. Compared to the Tesserae, our dataset is focused on nurses instead. Our dataset is much smaller in scope but focuses on stress bio-metrics. We contextualize this dataset as a unique data collection during the first year of COVID when nursing was stressed like never before. We use EDA and skin temperature in addition to the more common sensors of contemporary wrist-worn wearables such as pulse rate and accelerometer data that were not covered by either the TILES-2018 dataset or the Tesserae project.

[1]University of Louisiana at Lafayette, Lafayette, LA, USA. [2]Opelousas General Health System, Opelousas, LA, USA. ✉e-mail: Raju.Gottumukkala@louisiana.edu

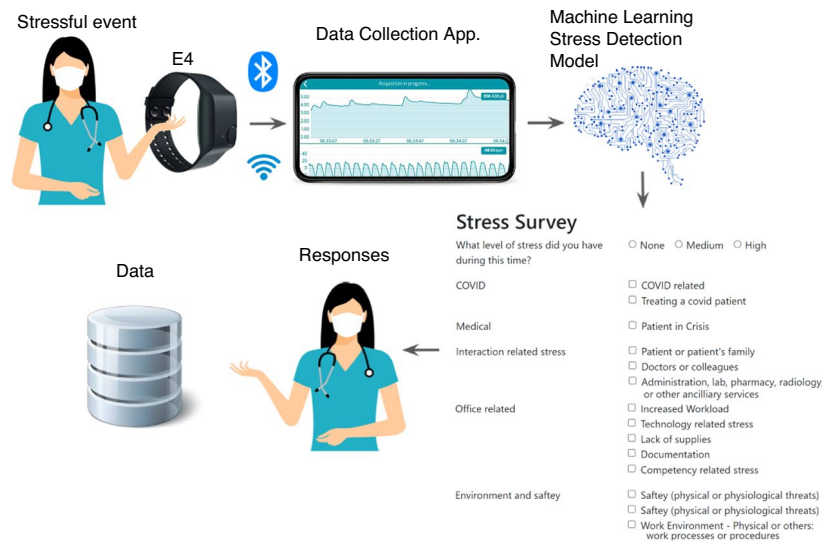

**Fig. 1** Stress detection apparatus.

Our dataset provides physiological stress signals collected using signal streams from Empatica E4 for a nursing population. Our primary motivation to create this dataset was to conduct a stress study under real-world work conditions. Further, this study was conducted during the COVID-19 outbreak. Nursing is a stressful profession, and prior literature identified several factors that contribute to stress. Moreover, the study was conducted during the second wave of the COVID-19 outbreak, and all the nurses were dealing with the influx of COVID-19 patients, which made it an event-rich environment. The combination of wearable data and end-of-shift surveys offers a useful window into nursing stress.

## Methods

In this section, we describe the experimental procedure and materials used for data collection. Fig. 1 shows the overall stress detection apparatus used for data collection. The Institutional Review Board of the University of Louisiana at Lafayette approved the protocol of the study: FA19–50 INFOR.

**Experimental procedure.** *Recruitment.* The research team presented the overall study design and approach to measure stress to the executive of the hospital, nurse managers, and human resources. After the nursing department expressed interest, the research team obtained approval from hospital compliance. The consenting nurse subjects were enrolled in the study. The stress detection scenario was described to the nurses from the Emergency Room (ER) department. The research team also ran a pilot of the study for a week and adjusted the frequency of survey responses to minimize inconvenience to the nurse's job duties. The overall study was done in three phases, where 7 nurses were included in each phase. No incentives were offered. Six nurses, however, did not complete the study, and these participants were dropped from the dataset. The subjects consented for the data to be publicly released with anonymization. Each subject was assigned a unique identifier that cannot be linked to the participant.

*Duration and demographics.* Data was gathered for approximately one week from 15 female nurses working regular shifts at a hospital. The age of the nurses ranged from 30 to 55 years. This amounts to 1,250 hours of data collected between 04/14/2020 to 12/11/2020 in two phases (Phase-I is from 4/15/2020 to 8/6/2020 and Phase-II from 10/08/2020 to 12/11/2020). The exclusion criteria were pregnancy, heavy smoking, mental disorders, and chronic or cardiovascular diseases. Table 1 presents the signals captured from individual participants that can be used for stress analysis.

*Data collection.* The subjects recruited in the study were nurses working on their usual schedules. Wearable devices can be worn during regular shifts and continuously monitor the physiological signals with minimum intrusion. In this study, we aimed to detect the stress of nurses during their daily routine using only wearables. An Empatica E4 was worn on the wrist of the dominant arm. We instructed our subjects to keep the phone in proximity and wear the device to maintain close contact with the skin to minimize data loss. We only include the data where the signals are collected continuously throughout the nurse's work shift (which is typically 8 hours). The nurse can terminate the data collection by turning off the device. In some cases, the collected data does not include Inter-Beat Interval (IBI) and Heart Rate data due to noisy artifacts from the PPG sensor. We did not notice any missing data from the EDA sensor. If any of the data collected during the shift is missing, the data from that shift is removed.

| Signal | Abbreviation | Frequency |
|---|---|---|
| electrodermal activity | EDA | 4.0 Hz |
| Heart Rate | HR | 1.0 Hz |
| skin temperature | ST | 1.0 Hz |
| accelerometer | ACC | 32 Hz |
| inter-beat interval | IBI | 64 Hz |
| blood volume pulse | BVP | 64 Hz |

**Table 1.** Signals and frequency of Empatica E4.

| ID | Total Data Collected | Duration of Stress Detected | Number of Stress Events Detected | Feedback (Agreement) | | Feedback Not Received |
|---|---|---|---|---|---|---|
| | | | | Yes | No | |
| 15 | 72:50 | 9:28 | 30 | 16 | 2 | 12 |
| 83 | 149:47 | 15 | 30 | 16 | 5 | 9 |
| 94 | 80:30 | 14:06 | 43 | 9 | 11 | 23 |
| 5C | 99:19 | 11:48 | 15 | 10 | 2 | 3 |
| 6B | 58:06 | 14:42 | 23 | 11 | 2 | 10 |
| 6D | 23:57 | 4:08 | 4 | 3 | 1 | 0 |
| 7A | 79:18 | 16:52 | 47 | 32 | 3 | 12 |
| 7E | 50:42 | 2:12 | 7 | 3 | 4 | 0 |
| 8B | 27:14 | 4:41 | 17 | 13 | 3 | 1 |
| BG | 76:02 | 8:28 | 25 | 14 | 4 | 7 |
| CE | 111:01 | 16:12 | 20 | 7 | 0 | 13 |
| DF | 127.41 | 17:27 | 21 | 6 | 2 | 13 |
| E4 | 116.12 | 17:56 | 40 | 29 | 6 | 5 |
| EG | 107.52 | 8:07 | 11 | 5 | 0 | 6 |
| F5 | 69:38 | 4:39 | 26 | 25 | 1 | 0 |

**Table 2.** Stress events and feedback.

During these experiments, we collected the physiological data from the nurses continuously from the start of their shift to the end of the shift. The overall end-to-end system was designed to perform data collection, processing, and stress detection in near real-time. While the proposed data collection mechanism, stress detection algorithm, and the survey apparatus were designed for real-time stress detection and feedback collection. The stress notifications were sent, and the survey responses were collected at the end of the shift for each of the nurses. This process reduces interruptions for nurses while on duty. While the delay may have produced some degree of recall bias, it is still an improvement over traditional surveys, which do not specify the precise time of stress events and are conducted at monthly or quarterly intervals. Data is transmitted from the wearable to the phone using Bluetooth, and the phone transmits the data to the analytics server using Wi-Fi in the background without the involvement of the participant. The stress detection algorithm uses a Random Forest model to identify epochs of potential stress. The nurses responded with a survey at the end of the day from their mobile phones, where they validated the detected events. 171 hours of data were detected as stressful. Table 2 shows the data description. The data was anonymized to remove publicly identifiable information and is available on Dryad.

**Data collection tools.** Figure 1 presents the data collection framework. A detailed explanation of each of the components is presented below.

*Empatica wristband.* An E4 wristband device (Empatica Inc., Milano, Italy) that collects physiological data such as EDA, Heart Rate, skin temperature, and accelerometer data from the right wrist of the subject. EDA is measured via E4's silver (Ag) electrode (valid range [0.01–100] $\mu$ S), while Heart Rate is measured via E4's PPG sensor. The E4 wristband is powered by a rechargeable lithium battery and transmits data to the subject's smartphone, using Bluetooth, in near-real-time. All the data collected from the E4 wristband and the sampling frequencies are presented in Table 1. The physiological data is then transmitted to the data collection app on a nurse's phone in near real-time. The nurses can also tag the data using the tag button on the e4 device to indicate an undetected stress experience, and this is also transmitted to the data collection app.

Nurses accessed the survey instrument through their mobile phone to validate the stress response. The survey instrument has the list of events detected (by the model) along with the time these events occurred. The survey instrument allows them to provide context for the event. Because some events may not have been detected, the nurses were asked to report any stress events undetected by the model through the survey instrument. The nurses also had an additional mechanism to report the events from the wristwatch by pushing the E4 button to tag stress events. Although, no one used the watch to report additional stress events.

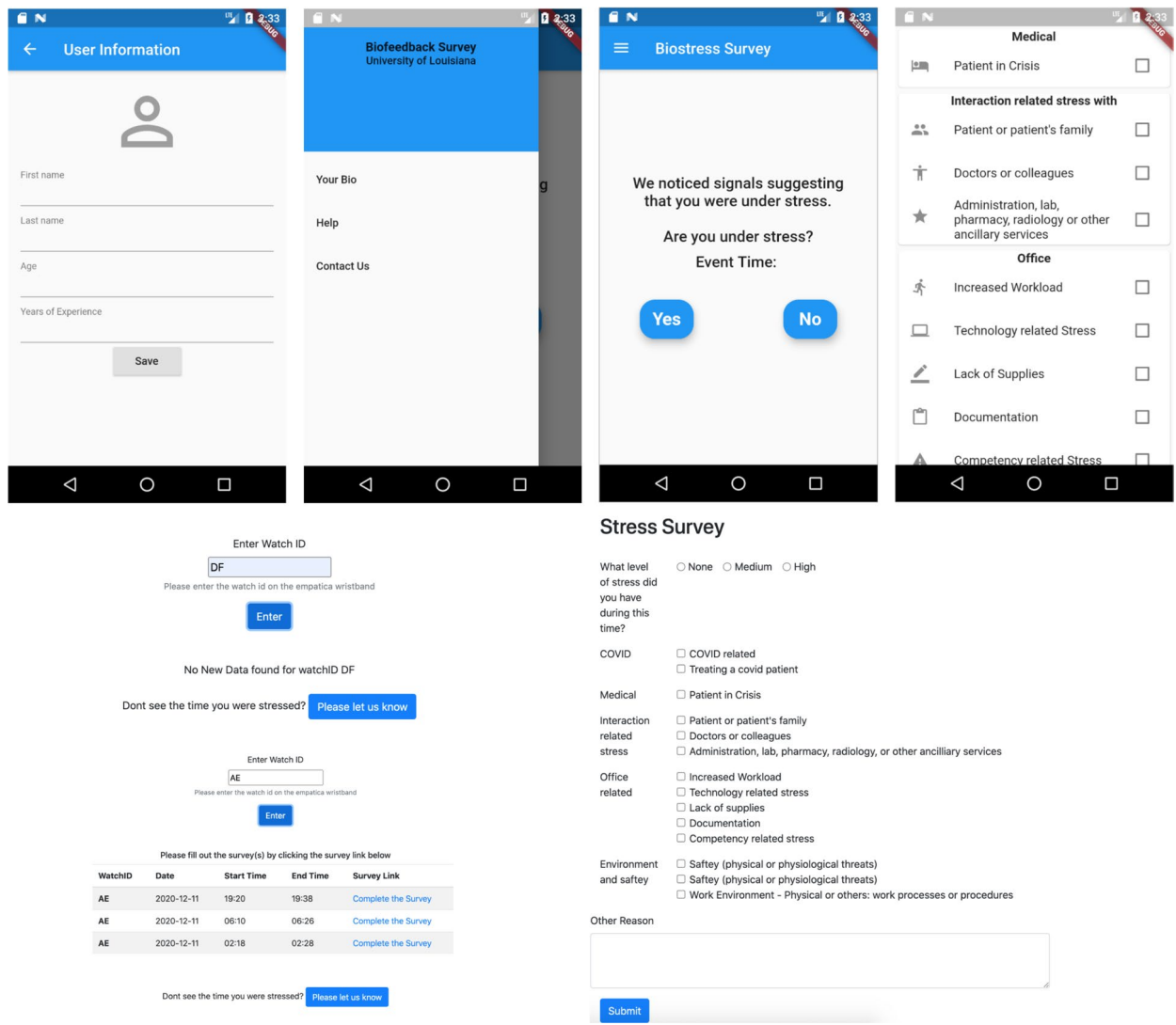

**Fig. 2** Mobile and Web application screenshots.

*Data collection application.* The data collection application is a mobile application that runs on iOS and Android. It connects to the E4 wristband through Bluetooth. The physiological signals are collected in near real-time. These signals and any tags input by the nurses are transmitted to the machine learning based stress detection model. Fig. 2 shows screenshots of the data collection app.

The survey instrument was designed for collecting survey responses. We offered nurses two ways of completing the surveys after stress detection: a custom mobile application and a web application (both offered the same functionality). The installation of the mobile application on the phone proved to be inconvenient as the application was not available on the App store and needed to be sideloaded onto the subject's phone. So, the nurses opted to fill out the survey through the web application. The source code of the custom applications is made available on GitHub[20].

The data collected on the watch is transmitted to the Empatica application on the subject's phone through Bluetooth. These signals are then transmitted to the Empatica server through the mobile phone's Wi-Fi connection. In the event of a loss of network connectivity, data is buffered (stored temporarily) on the phone and uploaded when the connection is restored. The researcher downloads data from the Empatica server to inspect the data for any losses to run the machine learning model.

*Machine learning and stress detection model.* At the end of the working day, the machine learning based stress detection model consumes the physiological and accelerometer data gathered from the E4 wristbands. Stress signals are detected from these physiological signals over time using a stress detection algorithm. The skin temperature and lag-based features were used in the stress detection model. In cases where the model detects several stress incidents, the nurses were asked to label the six longest durations per shift. The nurses had two ways to enter the labels. Nurses can select and edit the stress events detected by the stress detection system at various time slots and fill out a survey to indicate if they experienced stress, and if they did, the stress level, and the contributors of stress. They also had the option to report additional time slots where they experienced stress, along with the stress level and contributors of stress.

| Category | Stress inducer |
|---|---|
| COVID | COVID related [CR] |
|  | Treating a COVID patient [TCP] |
| Medical | Patient in crisis [PiC] |
| Interaction related stress | Patient or patient's family [PoPF] |
|  | Doctors or colleagues [DoC] |
|  | Administration, lab, pharmacy, radiology, or other ancillary services [Ad] |
| Office-related stress | Increased Workload [IWL] |
|  | Technology related stress [TR] |
|  | Lack of supplies [LoS] |
|  | Documentation [Doc] |
|  | Competency related stress [CRS] |
| Environment and safety | Safety (physical or physiological threats) [Saf] |
|  | Work Environment - Physical or others: work processes or procedures [WE] |

**Table 3.** Survey questions and their categories.

The machine learning based stress detection model was run on the physiological data (collected from the E4 wristbands at the end of each shift). The events detected by the model were transmitted back to the participant as part of the survey instrument. More details about the survey instrument are provided in the Survey sub-section below.

*Stress detection algorithm.* We trained, tested, and validated the Random Forest based machine learning model on the Affective Road dataset. Three signals were used, namely EDA, Skin temperature, and Heart Rate. This trained model was used for stress detection in nurses.

The stress detection algorithm has three parts. First, a sliding window of 10 seconds and a step-size of 5 seconds were used to extract signal features. The features used by the model include statistical features (mean, min, max, skewness, kurtosis, number of peaks) of EDA, skin temperature, and Heart Rate for the current window. We include three features (mean of skin temperature, Heart Rate, and EDA) from the previous 10 windows to provide an antecedent context of signals before stress. This resulted in 48 features. In the second stage, these 48 features are fed into the pre-trained machine learning algorithm to generate labels that represent stress categories (represented 0, 1, and 2 where 0 = no stress; 1 = low stress; 2 = high stress). Random Forest model was trained on the AffectiveRoad dataset, hyper-parameters were optimized using a grid search across a range of valid parameters: number of estimators (300, 400, **500**, 700, and 1000), minimum samples per leaf (3, 4, **5**, 6, and 7), maximum depth (10, 12, **15**, 18, 20), and maximum number of features (auto, sqrt, and **log2**), and 10-fold cross-validation is used to ensure that the model is not overfitting. Finally, a sliding window-based change point detection algorithm from Ruptures package[21] was used to identify discrete sessions from continuous stress signals. A sliding window based segmentation is used to determine a set of change points[22]. Each session is represented by a start time, end time, and stress label, where the label is determined by the average stress value 'S' between the start time and end time. The labels for each session are calculated based on the average stress values during the session: 'no stress' if S<=0.65, 'medium stress' if 0.65<=S<=1.3, and 'high stress' if S = 'high stress' S>=1.3. These thresholds were adopted based on the AffectiveRoad datasets[5] and were further validated using the survey feedback during the initial phase of the study. The source code for the stress detection algorithm comprises of feature extraction, stress detection, and change-point detection is provided in the GitHub repository[20].

*Survey.* The nurses were requested to complete a questionnaire about their experience during the detected stress events to identify the cause of the stress. The survey itself was not administered during the event in order to not add to the stress. Instead, the survey was administered at the end of the shift. Our approach to conduct surveys after a relatively short duration after the stress event (i.e., immediately after the shift) helps minimize the recall bias[23] associated with traditional surveys, which are administered after much longer durations such as a month[24] or a quarter. Our original research design before the pandemic had alerts sent to nurses an hour after the stress event had subsided. However, with the increased workload during the pandemic, it was suggested that we limit surveys to minimize disruptions. Conventional surveys also do not target specific stress events and are instead based on general recall of events across the epoch.

The questions in the questionnaire were selected based on a review of literature studying stress on nurses in a hospital environment, as well as from our discussions with nurses[25–36]. A list of questions in the survey is presented in Table 3.

| Filename | Columns | Measure | Description | Unit |
|---|---|---|---|---|
| ACC.csv | Column I | Accelerometer x-axis | Acceleration of the device along the x-axis | $m/s^2$ |
| | Column II | Accelerometer y-axis | Acceleration of the device along the y-axis | $m/s^2$ |
| | Column III | Accelerometer z-axis | Acceleration of the device along the z-axis | $m/s^2$ |
| BVP.csv | Column I | BVP | The volume of blood that passes through the tissues in the wrist and is used to measure IBI and Heart Rate | N/A |
| HR.csv | Column I | Heart Rate | A derived metric that measures the number of beats per minute based on Blood Volume Pulse | bpm |
| EDA.csv | Column I | EDA | Measurement of the skin conductivity levels | $\mu S$ |
| IBI.csv | Column I | Time | Time interval | Second |
| | Column II | IBI | Beat-to-beat interval | Second |
| TEMP.csv | Column I | Skin Temperature | The external temperature of the skin | Celsius |

**Table 4.** Empatica E4 Signal Description.

| Feature | Entropy | Information Gain |
|---|---|---|
| **EDA** | | |
| EDA_Mean | 0.126 | 0.102 |
| EDA_Min | 0.112 | 0.219 |
| EDA_Max | 0.122 | 0.218 |
| EDA_Std | 0.053 | 0.018 |
| EDA_Kurtosis | 0.014 | 0 |
| EDA_Skew | 0.017 | 0.002 |
| EDA_Num_Peaks | 0.003 | 0 |
| EDA_Amplitude | 0.016 | 0.035 |
| EDA_Duration | 0.009 | 0.015 |
| **Heart Rate** | | |
| HR_Mean | 0.041 | 0.005 |
| HR_Min | 0.042 | 0.009 |
| HR_Max | 0.043 | 0.01 |
| HR_Std | 0.019 | 0.002 |
| HR_RMS | 0.023 | 0.002 |
| **Skin Temperature** | | |
| temp_Mean | 0.111 | 0.056 |
| temp_Min | 0.096 | 0.036 |
| temp_Max | 0.105 | 0.04 |
| temp_Std | 0.041 | 0.012 |

**Table 5.** Entropy and Information gain of different features.

## Data Records

The complete dataset is available on Dryad[37]. The dataset contains 15 folders of physiological signals extracted from Empatica E4 wristbands for each participant. The corresponding stress survey responses from 15 nurses who wore the wristband is provided in SurveyResults.xlsx file. Both these files are linked by the Nurse identifier and the date-time field.

**Preprocessing.** The four signals: Heart Rate, skin temperature, EDA, and BVP, have different sampling rates. The frequency of these signals ranges from 1 Hz for the Heart Rate to 64Hz for the BVP. The frequency of EDA and skin temperature range from 4 Hz and 10 Hz each. For our stress detection, we use a frequency of 4 Hz to minimize information loss; this provided higher accuracy with a reasonable computation time on the pre-trained data that was used to predict the stress level. We evaluated various frequencies to detect stressful events using physiological stress. Our analysis shows that the computation time increases as the sampling frequency of the signals increase. In addition, certain signals like Heart Rate with lower sampling frequency are ignored due to low variability in the signal. Given the differences in the frequency of the signals from E4, we need to select a single frequency rate to process the signals. We evaluated different scenarios with various window sizes and frequencies for stress detection using the AffectiveRoad dataset. We observed that as the window size increases, the accuracy of stress detection decreases. Similarly, the accuracy is higher for higher frequencies. However, higher frequencies also result in an increase in computation time. Therefore a frequency of 4Hz is selected to maintain a balance between higher accuracy and reasonably near real-time stress detection. The data points for the three signals are interpolated to 4 Hz linearly. The code associated with the signal interpolation, cleaning, and pre-processing is provided along with the code repository. In addition, the raw signals with their original frequencies from Empatica E4 are also provided.

**Data description.** The following is a description of various directories and files in the dataset.

**Stress_dataset.zip**: The zip file holds the data of 15 participants in different folders. Each folder contains raw data signals in CSV format in a sub-folder. A raw data folder consists of 6 different CSV files, including (1) EDA. csv (EDA), (2) HR.csv (Heart Rate), (3) TEMP.csv (skin temperature), (4) IBI.csv (IBI), (5) BVP.csv (BVP), and (6) ACC.csv (accelerometer data). Each biometric signal data has the following information:

- **Start time (epoch):** The DateTime floating point value that contains the time that signal was generated using the internal clock of the wristband. The DateTime is stored in the first row of every data column.
- **Frequency:** The second cell of each column shows the data collection frequency (32)

Table 4 shows the csv files, column descriptions and units.

Each file name is identical to the participant's ID in both data and survey files. All of the signals were synchronized to bring them to a common frequency. The accelerometer data is not used in the stress detection model. Some of the basic physical activities can be estimated from the accelerometer sensor, which could be further used to potentially include the activity context in stress detection[38].

**SurveyResults.xlsx**: The Excel file holds all participant survey results and their annotated stress level in 11 Excel sheets (a sheet for each participant). Sheet names are the participant's IDs. However, the IDs are generated in an ID column for all files for more convenience. The following are the excel sheet columns:

**Group 1**: General information of the stress event.

- **Column A: ID** Anonymized Id of the user.
- **Column B: Start time:** Event start time.
- **Column C: End time:** Event start time.
- **Column D: Duration** Duration of the event.
- **Column E: Date** Date of data collection.

**Group 2**: Stress Level.

- **Column F: Stress level** Reported stress level by the nurse.

**Group 3:** Nurses' responses regarding the nature of the stress.

- **Column G: COVID Related**
- **Column H: Treating a COVID patient**
- **Column I: Patient in Crisis**
- **Column J: Patient or patient's family**
- **Column K: Doctors or colleagues**
- **Column L: Administration, lab, pharmacy, radiology, or other ancillary services**
- **Column M: Increased Workload**
- **Column N: Technology related stress**
- **Column O: Lack of supplies**
- **Column P: Documentation**
- **Column Q: Safety (physical or physiological threats)**
- **Column R: Lack of supplies**
- **Column S: Work Environment - Physical or others: work processes or procedures**
- **Column T: Description**

We consulted with the nurses ahead of the study to ensure that the stress labels were meaningful to them. The description field was used in place of "None of the Above". If the nurses agreed that there was stress and it was not covered by any stress label, it would be used to describe the stress event. It should be noted that it was also used to elaborate on the stress event even when one of the stress classes was selected.

## Technical Validation

Methods for evaluating psychological stress detection include self-report questionnaires and interviews. While stress surveys are considered sufficiently reliable and widely adopted[39], they offer insights mainly on the moments of their administration. The responses are coarse designations of stress and are unable to detect subtler shifts in stress over time[40]. In this paper, we used near-real-time stress evaluation surveys at the end of the shift in conjunction with our bio-metric stress detection to minimize the biases of recall.

**Label description.** The dataset provides more than 1,250 hours of accelerometer and physiological signals collected from 15 nurses during their daily routine responsibilities. 83 hours of data are labeled with stress descriptors based on the validated stressful events by nurses. We included the unlabeled signals since we expect the unlabeled signals to have predictive value in anticipating future stress events. Table 2 shows the stress detection results of participants based on their answers. Fig. 3 shows the distribution of stress events in the nurses. Table 2 shows the data durations we collected from each participant.

While the nurses were instructed to use the E4 buttons to indicate the situations where they were stressed, none of the nurses actually used them. The false positives are identified based on the surveys of stress events where the nurses suggested that they were not stressed during the stress event detected by the stress detection model. The nurses were also provided the opportunity to provide additional times when they were stressed but were not detected by the model. But no additional data points were provided by the nurses beyond what was detected by the stress detection model.

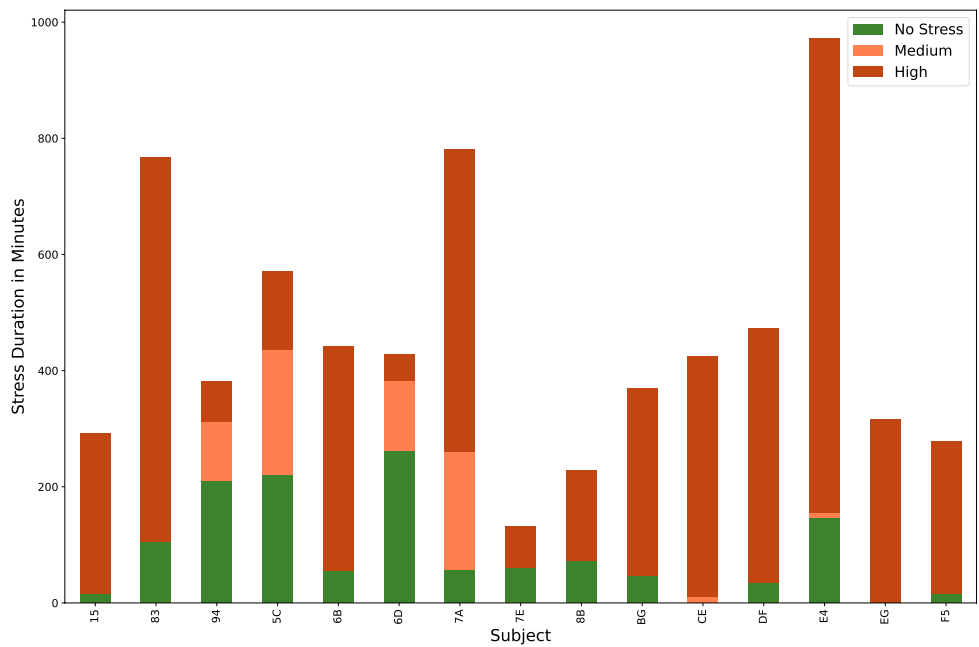

**Fig. 3** Distribution of stress levels for each subject.

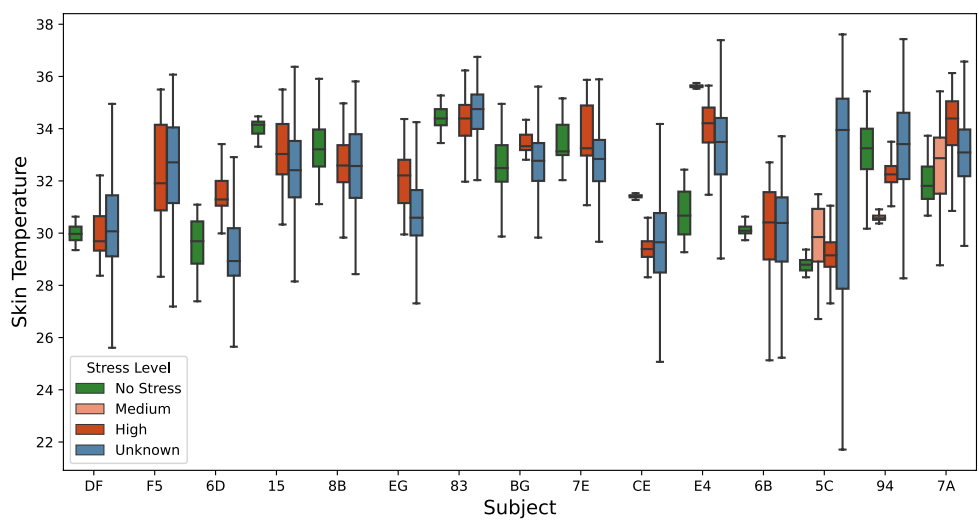

**Fig. 4** Overall skin temperatures of participants.

The "na" events are labelled using the stress detection model but are not validated by the subject, unlike the stress events, which are validated. We requested the subjects to add undetected stress events at the end of the survey. This would validate the No Stress epochs, but we did not receive responses. So, only the labels 0, 1, and 2, corresponding to no-stress, medium-stress, and high-stress, respectively, are validated.

**Physiological signals versus reported stress events.** The box plots (Figs. 5–8) show four individual physiological signals recorded by the wearable device, namely the Heart Rate, EDA, skin temperature, BVP, and the reported stress levels from the survey. Pekka et al.[41], have shown that the stress detection models are personalized signals and features, and their importance in the machine learning models has to be computed at the user level. Given the same signals, feature selection from within the signal streams at the user level improves performance, as demonstrated by Pekka's 5-fold cross-validations. Figs. 4, 5, 6 illustrate that the biometric signals like EDA, Heart Rate, and skin temperature vary between subjects.

The relationship between these signals and stress is not necessarily linear. In addition, there is also an interplay between these signals. Machine learning techniques can model stress behavior with respect to each of these features derived from the signals. In Table 5, we provide the entropy and information gain for the random forest model to identify the stress level of the subjects. The entropy is higher for EDA-based signals compared to skin temperature and Heart Rate. The EDA-based features (mean, min, and max) for each of the subjects within a given window is a better predictor of an individual under stress. While earlier studies indicated that Heart Rate and Heart Rate variability are good features to use for stress detection[42,43], this is not evident in our analysis. This could be due to various reasons. Earlier studies relied on simulated data in laboratory settings where the subjects were in stationary/sedentary positions. However, in our real-world analysis, there is a lot of physical activity performed by nurses, leading to their Heart Rates being elevated.

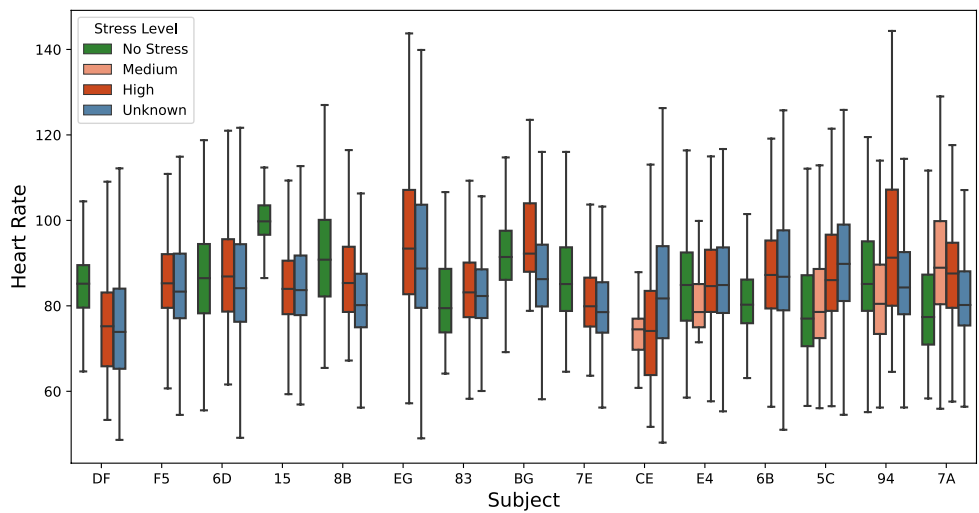

**Fig. 5** Overall HR of participants.

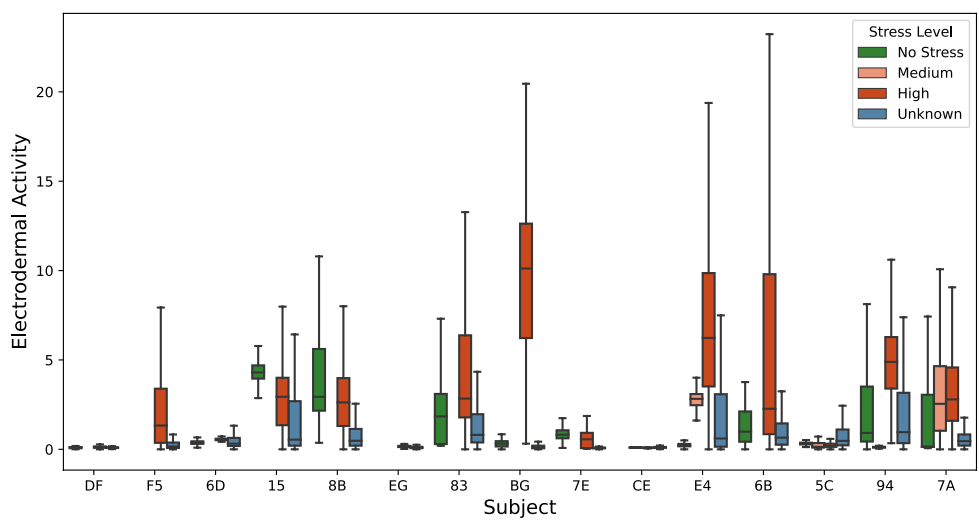

**Fig. 6** Overall EDA of participants.

The skin temperature of a healthy individual is about 33°C or 91°F. Skin temperature varies during various activities due to skin blood temperature and its flow and is normally within the range of 33.5°C to 36.9°C[44]. However, this can vary quite widely based on the type and length of activity and room temperature. Given the open-ended nature of the experiment, there are some anomalies in the data. The subjects were twice as likely to have a higher skin temperature with high stress than during no stress. The general trend agrees with reports from prior literature. However, the trend was not statistically significant in our dataset. We analyzed the data with a paired sample t-test. With the test statistic of 1.3 and the p-value at 0.22, we cannot reject the null hypothesis that there is no difference in temperature with respect to stress[45]. The relationship between skin temperature and stress has been discussed by several authors[45,46] and observed skin warming in stressful events.

The Heart Rate of a healthy individual, irrespective of gender, ranges from 60 to 100, when in a resting state. However, the Heart Rate varies greatly with activity. Given that the subjects are typically performing different activities, one can observe high variations in Heart Rate. Fig. 5 shows the distribution of Heart Rate and associated stress level for all the subjects. Based on the visual inspection of the plot, stress does not appear to have a strong correlation with the Heart Rate. Heart rate is generally associated with high stress situations, but high Heart Rate should not imply high stress since it is more commonly influenced by non-stressful physical activity.

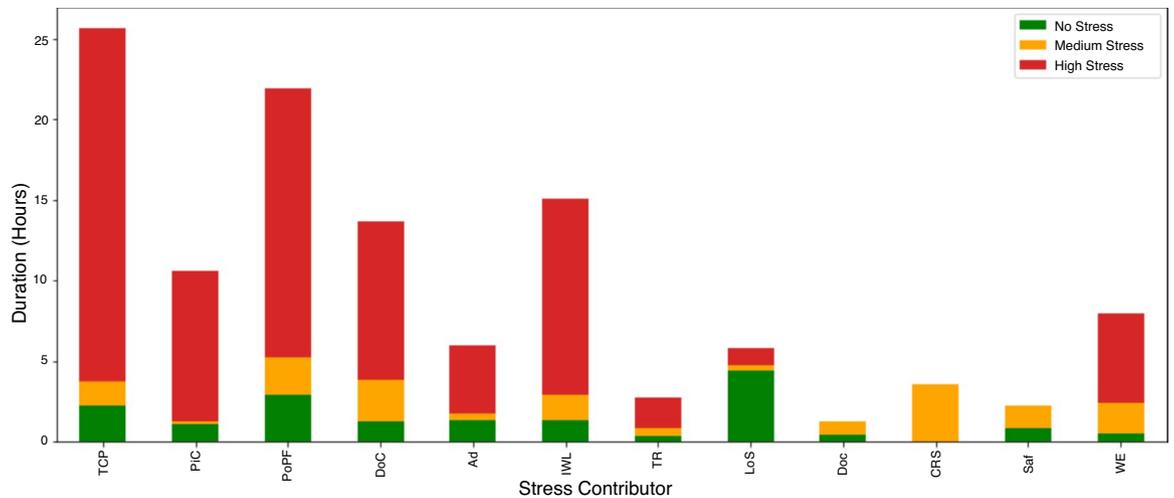

**Fig. 7** Overall BVP of participants.

Heart rate is best interpreted with accelerometer data, which can signify physical activity. We do not currently provide activity recognition labels and models from wrist-based accelerometer data.

Figure 6 shows the distribution of EDA and associated stress levels for all the subjects. Based on visual inspection of the plot, stress has a positive correlation with the EDA. The average EDA is higher in stressful situations for some participants. However, for some participants, the EDA is not a good indicator of stress because EDA does not vary or it is not positively correlated. There is high variability in EDA signals for various subjects in stressful events. The normal range for humans is from 1 to 20 $\mu S$. We observe that the average skin EDA for all the participants when there is no stress reported is below 5, and the range for medium stress is the same as stress-free situations. The EDA contains two main components, skin conductance response (SCR) and skin conductance level (SCL). The skin conductance response is a phasic response to external stimuli, whereas SCL is a gradual change in skin conductance over time. In our analysis, the SCL plays a major role, as the stress detection is performed in windows and compared to a normal baseline. For example, the EDA was observed to reach a maximum of 60 $\mu S$ compared to 10 $\mu S$ during non-stress time-periods on average.

The Heart Rate and Heart Rate variability signals can be derived from the BVP signal by computing the inverse of the time between two successive peaks. Fig. 7 shows the overall distribution of the BVP signal of different participants. We analyzed the BVP signals using a paired sample t-test with the test statistic of 2.3 and the p-value at 0.0501. We cannot reject the null hypothesis that there is no difference in BVP with respect to stress.

Figure 8 shows the distribution of stress intensity within each of the contributors to stress, as well as the cumulative number of hours under each stress level per contributor. Table 3 provides additional details about the factors. The survey results indicate that treating a COVID-19 patient was the most significant contributor to detected high-stress events. Not all COVID-related stress classes ranked high, however. There was no event where the nurses are worried about contracting COVID. In addition to COVID, the nurses indicate that they were impacted by the lack of supplies; however, it was not mentioned as a contributor of acute stress for any of our detected events. This could mean that while COVID-19 and supplies could have been significant contributors to stress in the background, they themselves did not cause acute stress events detectable by our apparatus. These results are also verified by other researchers studying the impact of COVID on nurses[47]. This should be seen as a limitation of biometric stress detection; it cannot detect general concerns, and in a holistic stress evaluation, biometric data collection must always be paired with qualitative interviews. It is a complementary modality, rather than a replacement, to traditional stress monitoring.

### Usage Notes

**Potential applications.** Human well-being is an important consideration both for individuals and organizations. As such, organizations need mechanisms to carefully monitor and manage high-stress environments such as hospitals in order to improve both employee wellbeing as well as patient satisfaction. We believe this study can be useful for researchers from many domains. First, researchers in signal processing and machine learning might be able to use the dataset to develop new machine learning models that improve stress detection performance. Second, the accelerometer signals can be used to contextualize stress in order to understand the relationship between movement patterns and stress.

The data collected here can be useful for nursing, machine learning, and hospital management communities. The physiological signals, along with the activity data, can be used to better contextualize signal streams that are impacted by the intensity of physical activity, such as Heart Rate. The interpretation context of our dataset is significantly different given that our subjects were engaged in real-world tasks that have varied physical components to them. Thus, further analyses can benefit from augmentation with inferred metrics of mobility.

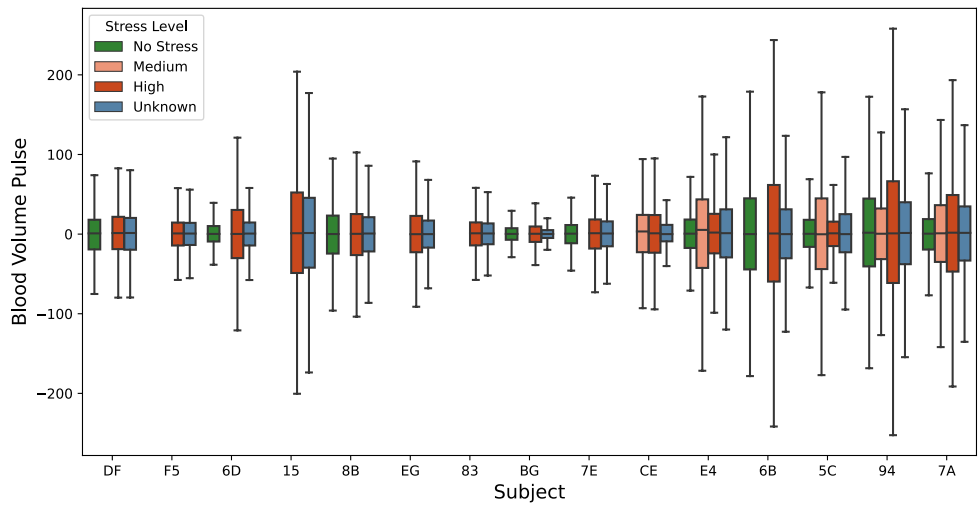

**Fig. 8** Distribution of stress levels within detected events across stress contributors.

Finally, researchers in human resources / human factors / organizational psychology would find the survey dataset, along with biometric signals useful because it is a unique dataset that makes the association between biometric signals and stress-related factors during the COVID-19 outbreak.

Additionally, the dataset illustrates the relative frequency of various work-related stressors during the pandemic.

**Limitations.** We conceptualized the study before the pandemic. The original design had an onsite observer making independent assessments of tasks and apparent stress behaviors alongside the system stress assessments. The outbreak prevented us from placing an investigator onsite but provided a dataset under critically unique clinical circumstances. The nurses were busier than usual, and we had to ensure that we were not interrupting them too often.

Not all of the dataset is covered by stress labels. Unlabelled data does not necessarily imply a lack of stress; it just means that we did not detect stress using our signal streams and that the subjects did not independently report it as a stressful period. We did not insist on complete coverage of labels, as the most important priority of nurses was taking care of the patients.

We also provide the unlabelled data because we suspect that it may contain predictive markers of stress that future analyses may reveal.

Our study is distinguished by the following features and limitations that must be taken into account while interpreting the data. These are simply inherent to the naturalistic setting and due to the mitigating effects of the pandemic scenario.

- Compared to the laboratory scenarios where the undivided attention of the subject is available, in high-impact real-world scenarios, the subjects may not be distracted or interrupted frequently from their professional tasks for labeling.
- We only validated stressful events because of our focus on acute stress detection. Because nursing was stressful during the peaks of the COVID-19 pandemic, we provided no more than 6 events a day for the nurses. When there are more than 6 stress events, we prioritized stress events of longer duration since we expected the subjects to recall these events better at the end of the day. We also chose to spread the events across the nurses' work shifts since this is more likely to reduce recall conflicts.
- In prior literature, researchers have claimed that chronic stress can be mitigated by early detection of acute stress[40,48]. The literature does not, however, validate this claim. There is not sufficient research on chronic stress detection using biometric signals[49]. This dataset is also focused on acute stress detection and is unable to detect chronic stress.
- Social distancing requirements upended a prior experimental design that included onsite investigator data collection describing nurse tasks in conjunction with stress signals.
- While we had methods for immediate stress detection, we opted to delay the surveys to the end of the day in order to not interrupt the work of the nurses. While the latency may have produced some degree of recall bias, it is still an improvement over traditional surveys, which do not specify the precise time of stress events and are conducted at monthly or quarterly intervals[24].
- The nurses agreed with the stress detection algorithm far more than they disagreed. Given the stress of the pandemic, and because the stress reports of nurses were not corroborated by onsite investigator observations of stressful behavior, the data can only be interpreted as subject reports of stress.

Finally, unlike laboratory studies that are typically conducted in a controlled environment, stress detection in a natural environment is more complex due to the influence of many social, cultural, and psychological factors.

While we have attempted to provide some context to stress by conducting a survey based on the available literature, there are several social, cultural, and individual variables we did not consider in our survey. Given the stress of the pandemic, and because the stress reports of nurses were not corroborated by onsite investigator observations of stressful behavior. The data should specifically be interpreted as subject reports of stress, which may be mitigated by other factors such as a desire to quickly complete the survey questions after a stressful day.

### Code availability

The code is available on GitHub[20]. The data collection was performed in Central Standard Time of the United States which is 6 hours behind GMT.

### References

1. Peternel, K., Pogačnik, M., Tavčar, R. & Kos, A. A presence-based context-aware chronic stress recognition system. *Sensors* **12**, 15888–15906 (2012).
2. Bickford, M. Stress in the workplace: A general overview of the causes, the effects, and the solutions. *Canadian Mental Health Association Newfoundland and Labrador Division* **8**, 1–3 (2005).
3. Wellen, K. E. *et al*. Inflammation, stress, and diabetes. *The Journal of clinical investigation* **115**, 1111–1119 (2005).
4. Greenglass, E. R., Burke, R. J. & Fiksenbaum, L. Workload and burnout in nurses. *Journal of community & applied social psychology* **11**, 211–215 (2001).
5. Haouij, N. E., Poggi, J.-M., Sevestre-Ghalila, S., Ghozi, R. & Jaïdane, M. Affectiveroad system and database to assess driver's attention. In *Proceedings of the 33rd Annual ACM Symposium on Applied Computing*, 800–803 (2018).
6. Schmidt, P., Reiss, A., Duerichen, R., Marberger, C. & Van Laerhoven, K. Introducing wesad, a multimodal dataset for wearable stress and affect detection. In *Proceedings of the 20th ACM international conference on multimodal interaction*, 400–408 (2018).
7. Kirschbaum, C., Pirke, K.-M. & Hellhammer, D. H. The 'trier social stress test'–a tool for investigating psychobiological stress responses in a laboratory setting. *Neuropsychobiology* **28**, 76–81 (1993).
8. Sriramprakash, S., Prasanna, V. D. & Murthy, O. R. Stress detection in working people. *Procedia computer science* **115**, 359–366 (2017).
9. Chen, W., Zheng, S. & Sun, X. Introducing mdpsd, a multimodal dataset for psychological stress detection. In *Big Data: 8th CCF Conference, BigData 2020, Chongqing, China, October 22–24, 2020, Revised Selected Papers*, **vol. 1320**, 59 (Springer Nature, 2021).
10. Birkett, M. A. The trier social stress test protocol for inducing psychological stress. *JoVE (Journal of Visualized Experiments)* e3238 (2011).
11. Scarpina, F. & Tagini, S. The stroop color and word test. *Frontiers in psychology* **8**, 557 (2017).
12. Mundnich, K. *et al*. Tiles-2018, a longitudinal physiologic and behavioral data set of hospital workers. *Scientific Data* **7**, 1–26 (2020).
13. Martinez, G. J. *et al*. Improved sleep detection through the fusion of phone agent and wearable data streams. In *2020 IEEE International Conference on Pervasive Computing and Communications Workshops (PerCom Workshops)*, 1–6 (IEEE, 2020).
14. Martinez, G. J. *et al*. On the quality of real-world wearable data in a longitudinal study of information workers. In *2020 IEEE International Conference on Pervasive Computing and Communications Workshops (PerCom Workshops)*, 1–6 (IEEE, 2020).
15. DATA, I. P. & EDUCATION, I. Curriculum vitae–aaron d. striegel. *Ethics* **16** (2004).
16. Mirjafari, S. *et al*. Differentiating higher and lower job performers in the workplace using mobile sensing. *Proceedings of the ACM on Interactive, Mobile, Wearable and Ubiquitous Technologies* **3**, 1–24 (2019).
17. Saha, K. *et al*. Imputing missing social media data stream in multisensor studies of human behavior. In *2019 8th International Conference on Affective Computing and Intelligent Interaction (ACII)*, 178–184 (IEEE, 2019).
18. Saha, K. *et al*. Social media as a passive sensor in longitudinal studies of human behavior and wellbeing. In *Extended Abstracts of the 2019 CHI Conference on Human Factors in Computing Systems*, 1–8 (2019).
19. Mattingly, S. M. *et al*. The tesserae project: Large-scale, longitudinal, in situ, multimodal sensing of information workers. In *Extended Abstracts of the 2019 CHI Conference on Human Factors in Computing Systems*, 1–8 (2019).
20. Ravi, M. S. Stress-detection-in-nurse. https://github.com/CPHSLab/Stress-Detection-in-Nurses (2021).
21. Truong, C., Oudre, L. & Vayatis, N. ruptures: change point detection in python. *arXiv preprint arXiv:1801.00826* (2018).
22. Truong, C., Oudre, L. & Vayatis, N. Selective review of offline change point detection methods. *Signal Processing* **167**, 107299 (2020).
23. Tarrant, M. A., Manfredo, M. J., Bayley, P. B. & Hess, R. Effects of recall bias and nonresponse bias on self-report estimates of angling participation. *North American Journal of Fisheries Management* **13**, 217–222 (1993).
24. Sveinsdottir, H., Biering, P. & Ramel, A. Occupational stress, job satisfaction, and working environment among icelandic nurses: a cross-sectional questionnaire survey. *International journal of nursing studies* **43**, 875–889 (2006).
25. Adriaenssens, J., De Gucht, V. & Maes, S. Causes and consequences of occupational stress in emergency nurses, a longitudinal study. *Journal of nursing management* **23**, 346–358 (2015).
26. Brown, S., Whichello, R. & Price, S. The impact of resiliency on nurse burnout: An integrative literature review. *Medsurg Nursing* **27**, 349 (2018).
27. Jovanov, E., Frith, K., Anderson, F., Milosevic, M. & Shrove, M. T. Real-time monitoring of occupational stress of nurses. In *2011 Annual International Conference of the IEEE Engineering in Medicine and Biology Society*, 3640–3643 (IEEE, 2011).
28. Gelsema, T. I., Van Der Doef, M., Maes, S., Akerboom, S. & Verhoeven, C. Job stress in the nursing profession: The influence of organizational and environmental conditions and job characteristics. *International Journal of Stress Management* **12**, 222 (2005).
29. Hersch, R. K. *et al*. Reducing nurses' stress: A randomized controlled trial of a web-based stress management program for nurses. *Applied nursing research* **32**, 18–25 (2016).
30. Khamisa, N., Oldenburg, B., Peltzer, K. & Ilic, D. Work related stress, burnout, job satisfaction and general health of nurses. *International journal of environmental research and public health* **12**, 652–666 (2015).
31. Kurki, R. Stress management among nurses: Literature review of causes and coping strategies (2018).
32. Kurnat-Thoma, E., Ganger, M., Peterson, K. & Channell, L. Reducing annual hospital and registered nurse staff turnover—a 10-element onboarding program intervention. *SAGE Open Nursing* **3**, 2377960817697712 (2017).
33. Lo, W.-Y., Chien, L.-Y., Hwang, F.-M., Huang, N. & Chiou, S.-T. From job stress to intention to leave among hospital nurses: A structural equation modelling approach. *Journal of advanced nursing* **74**, 677–688 (2018).
34. Lu, H., Zhao, Y. & While, A. Job satisfaction among hospital nurses: A literature review. *International journal of nursing studies* **94**, 21–31 (2019).
35. Gjoreski, M., Gjoreski, H., Luštrek, M. & Gams, M. Continuous stress detection using a wrist device: in laboratory and real life. In *proceedings of the 2016 ACM international joint conference on pervasive and ubiquitous computing: Adjunct*, 1185–1193 (2016).

4www.nature.com/scientificdata/www.nature.com/scientificdata/36. Lopez-Martinez, D., El-Haouij, N. & Picard, R. Detection of real-world driving-induced affective state using physiological signals and multi-view multi-task machine learning. In *2019 8th International Conference on Affective Computing and Intelligent Interaction Workshops and Demos (ACIIW)*, 356–361 (IEEE, 2019).
37. Hosseini, S. *et al.* A multi-modal sensor dataset for continuous stress detection of nurses in a hospital. *Dryad* https://doi.org/10.5061/dryad.5hqbzkh6f (2021).
38. Foerster, F., Smeja, M. & Fahrenberg, J. Detection of posture and motion by accelerometry: a validation study in ambulatory monitoring. *Computers in human behavior* **15**, 571–583 (1999).
39. Tsutsumi, A., Inoue, A. & Eguchi, H. How accurately does the brief job stress questionnaire identify workerswith or without potential psychological distress? *Journal of occupational health* 17–0011 (2017).
40. Alberdi, A., Aztiria, A. & Basarab, A. Towards an automatic early stress recognition system for office environments based on multimodal measurements: A review. *Journal of biomedical informatics* **59**, 49–75 (2016).
41. Siirtola, P. & Röning, J. Comparison of regression and classification models for user-independent and personal stress detection. *Sensors* **20**, 4402 (2020).
42. Melillo, P., Bracale, M. & Pecchia, L. Nonlinear heart rate variability features for real-life stress detection. case study: students under stress due to university examination. *Biomedical engineering online* **10**, 1–13 (2011).
43. Boonnithi, S. & Phongsuphap, S. Comparison of heart rate variability measures for mental stress detection. In *2011 Computing in Cardiology*, 85–88 (IEEE, 2011).
44. Tanda, G. The use of infrared thermography to detect the skin temperature response to physical activity. In *Journal of Physics: Conference Series*, vol. 655, 012062 (IOP Publishing, 2015).
45. Karthikeyan, P., Murugappan, M. & Yaacob, S. Descriptive analysis of skin temperature variability of sympathetic nervous system activity in stress. *Journal of Physical Therapy Science* **24**, 1341–1344 (2012).
46. Baker, L. M. & Taylor, W. M. The relationship under stress between changes in skin temperature, electrical skin resistance, and pulse rate. *Journal of experimental psychology* **48**, 361 (1954).
47. Morgantini, L. A. *et al.* Factors contributing to healthcare professional burnout during the covid-19 pandemic: a rapid turnaround global survey. *PloS one* **15**, e0238217 (2020).
48. Giannakakis, G. *et al.* Review on psychological stress detection using biosignals. *IEEE Transactions on Affective Computing* (2019).
49. Baumgartl, H., Fezer, E. & Buettner, R. Two-level classification of chronic stress using machine learning on resting-state eeg recordings (2020).
### Acknowledgements

This project is funded by NSF Grants 1650551, CNS-1429526, and by the Louisiana Board of Regents Support Fund contract LEQSF (2019–20)-ENH-DE-22. The authors would also like to acknowledge the reviewers for their constructive inputs that helped improve the clarity of the manuscript. The authors would also like to acknowledge the reviewers for their constructive inputs that helped improve the clarity of the manuscript.### Author contributions

Conceptualization: R.G., Z.A., K.C, and C.B.; Methodology: S.K., S.H., and R.B.; Software: S.K., S.H., and R.B.; Validation: S.K., S.H., and R.B, C.B., K.C., and R.G.; Formal analysis: S.H. S.K., and R.B.; Investigation: R.G., C.B., C.K., and Z.A.; Resources: R.G., and C.B.; Data duration: R.B. and S.K.; Writing—original draft preparation: S.H., and R.B.; Writing—review and editing: R.G., Z.A., C.B. and K.C.; Visualization: R.B., S.K. and S.H.; Supervision: R.G., Z.A., C.B. and K.C.; Project administration: R.G.; Funding acquisition: R.G., C.B., Z.A., and R.B. All authors reviewed the manuscript.

### Competing interests

The authors declare no competing interests.

### Additional information

**Correspondence** and requests for materials should be addressed to R.G.

**Reprints and permissions information** is available at www.nature.com/reprints.

**Publisher's note** Springer Nature remains neutral with regard to jurisdictional claims in published maps and institutional affiliations.

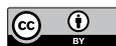

**Open Access** This article is licensed under a Creative Commons Attribution 4.0 International License, which permits use, sharing, adaptation, distribution and reproduction in any medium or format, as long as you give appropriate credit to the original author(s) and the source, provide a link to the Creative Commons license, and indicate if changes were made. The images or other third party material in this article are included in the article's Creative Commons license, unless indicated otherwise in a credit line to the material. If material is not included in the article's Creative Commons license and your intended use is not permitted by statutory regulation or exceeds the permitted use, you will need to obtain permission directly from the copyright holder. To view a copy of this license, visit http://creativecommons.org/licenses/by/4.0/.

© The Author(s) 2022SCIENTIFIC DATA |    (2022) 9:255  | https://doi.org/10.1038/s41597-022-01361-y    13